\documentclass[12pt]{article}
\usepackage{amssymb}
\usepackage{amsmath}
\usepackage[T1]{fontenc}
\usepackage{textcomp}

\usepackage{amsthm}
\begin{document}
\font\frak=eufm10 scaled\magstep1
\font\fak=eufm10 scaled\magstep2
\font\fk=eufm10 scaled\magstep3
\font\black=msbm10 scaled\magstep1
\font\bigblack=msbm10 scaled\magstep 2
\font\bbigblack=msbm10 scaled\magstep3
\font\scriptfrak=eufm10
\font\tenfrak=eufm10
\font\tenblack=msbm10

%A continuacion definimos los comandos para utilizar los
%fuentes en modo matematico

%Operadores especiales, abrev. matematicasxs
\def\biggoth #1{\hbox{{\fak #1}}}
\def\bbiggoth #1{\hbox{{\fk #1}}}
\def\sp #1{{{\cal #1}}}
\def\goth #1{\hbox{{\frak #1}}}
\def\scriptgoth #1{\hbox{{\scriptfrak #1}}}
\def\smallgoth #1{\hbox{{\tenfrak #1}}}
\def\smallfield #1{\hbox{{\tenblack #1}}}
\def\field #1{\hbox{{\black #1}}}
\def\bigfield #1{\hbox{{\bigblack #1}}}
\def\bbigfield #1{\hbox{{\bbigblack #1}}}
\def\Bbb #1{\hbox{{\black #1}}}
\def\v #1{\vert #1\vert}             %Para denotar elgrado de #1
\def\ord#1{\vert #1\vert}
\def\m #1 #2{(-1)^{{\v #1} {\v #2}}} %Para denotar el signo (-1)^...
\def\lie #1{{\sp L_{\!#1}}}               %%Lie derivative
\def\pd#1#2{\frac{\partial#1}{\partial#2}}
\def\pois#1#2{\{#1,#2\}}
 %  un parntesis de Poisson {f,g}
\def\set#1{\{\,#1\,\}}             %  notacin para conjuntos
\def\<#1>{\langle#1\rangle}        %  una forma bilineal <x,a>
\def\>#1{{\bf #1}}                %  notacin para vectores
\def\f(#1,#2){\frac{#1}{#2}}
\def\cociente #1#2{\frac{#1}{#2}}
\def\braket#1#2{\langle#1\mathbin\vert#2\rangle} %% <w|z>
\def\brakt#1#2{\langle#1\mathbin,#2\rangle}           %% <w,z>
\def\dd#1{\frac{\partial}{\partial#1}}
\def\bra #1{{\langle #1 |}}
\def\ket #1{{| #1 \rangle }}
\def\ddt#1{\frac{d #1}{dt}}
\def\dt2#1{\frac{d^2 #1}{dt^2}}
\def\matriz#1#2{\left( \begin{array}{#1} #2 \end{array}\right) }
\def\Eq#1{{\begin{equation} #1 \end{equation}}}

%%Abreviaturas de simbolos
\def\bw{{\bigwedge}}      %%from Marmo diff geom
\def\hut{{\scriptstyle \land}}            %%from Marmo diff geom
\def\dg{{\goth g^*}}                                                                                                            %%dual of the Lie algebra
\def\Cdg{{C^\infty (\goth g^*)}}
\def\poi{\{\:,\}}                           %parntesis de Poisson {,}
\def\qw{\hat\omega}                %  omega con sombrero
\def\FL{{\sp F}L}                 %  abreviatura para Transf. Legendre
\def\hFL{\widehat{{\sp F}L}}      %  abreviat. para Transf. Legendreext.
\def\XHMw{\goth X_H(M,\omega)}
\def\XLHMw{\goth X_{LH}(M,\omega)}
\def\ea{\varepsilon_a}
\def\ep{\varepsilon}
\def\mitad{\frac{1}{2}}
\def\x{\times}
\def\cinf{C^\infty}
\def\forms{\bigwedge}                 %  formas
\def\onda{\tilde}
\def\orb{{\sp O}}

%%letras griegas
\def\a{\alpha}
\def\d{\delta}
\def\g{{\gamma }}                  %  gama
\def\G{{\Gamma}}
\def\La{\Lambda}                   %  lambda
\def\la{\lambda}                   %  Lambda
\def\w{\omega}                     %  una forma simplectica
\def\W{{\Omega}}                   %  Omega
\def\ltimes{\bowtie}

%%letras cal, Bbb, goticas, etc.
\def\roc{{\tilde{\cal R}}}                       %%from Marmo diff geom
\def\cl{{\cal L}}                               %%from Marmo diff geom
\def\V{{\sp V}}                                 %espacio de velocidades
\def\F{{\sp F}}
\def\cv{{{\goth X}}}                    %  un campo vectorial
\def\LG{\goth g}
\def\LH{\goth h}
\def\X{{{\goth X}}}                     %  un campovectorial
\def\R{{\hbox{{\field R}}}}             %%real numbers (Pepin)
\def\big R{{\hbox{{\bigfield R}}}}
\def\bbig R{{\hbox{{\bbigfield R}}}}
\def\C{{\hbox{{\field C}}}}         %%complex numbers (Pepin)
\def\Z{{\hbox{{\field Z}}}}             %%real numbers (Pepin)
\def\N{{\hbox{{\field N}}}}         %%complex numbers (Pepin)
%\def\small C{{\hbox{{\smallfield C}}}}         %%smallcomplex numbers (Pepin)

%%notaciones rm en modo math
\def\ima{\hbox{{\rm Im}}}                               %%Image of a map
\def\dim{\hbox{{\rm dim}}}        %%several definitions
\def\End{\hbox{{\rm End}}}
\def\Tr{\hbox{{\rm Tr}}}
\def\tr{{\hbox{\rm\small{Tr}}}}                %%Trace
\def\lin{{\hbox{Lin}}}
\def\vol{{\hbox{vol}}}
\def\Hom{{\hbox{Hom}}}
\def\rank{{\hbox{rank}}}
\def\Ad{{\hbox{Ad}}}
\def\ad{{\hbox{ad}}}
\def\CoAd{{\hbox{CoAd}}}
\def\coad{{\hbox{coad}}}
\def\Rea{\hbox{Re}}                     %  parte real
\def\id{{\hbox{id}}}                    %  la identidad
\def\Id{{\hbox{Id}}}
\def\Int{{\hbox{Int}}}
\def\Ext{{\hbox{Ext}}}
\def\Aut{{\hbox{Aut}}}
\def\Card{{\hbox{Card}}}
\def\SODE{{\small{SODE }}}
\newcommand{\bea}{\begin{eqnarray}}
\newcommand{\eea}{\end{eqnarray}}

\def\R{\mathbb{R}}
\def\ba{\begin{eqnarray}}
\def\ea{\end{eqnarray}}
\def\be{\begin{equation}}
\def\ee{\end{equation}}
\def\Eq#1{{\begin{equation} #1 \end{equation}}}
\def\R{\Bbb R}
\def\C{\Bbb C}
\def\Z{\Bbb Z}
\def\a{\alpha}                  % alpha
\def\b{\beta}                   % beta
\def\g{\gamma}                  % gamma
\def\d{\delta}                  % delta
\def\bra#1{\langle#1|}
\def\ket#1{|#1\rangle}
\def\goth #1{\hbox{{\frak #1}}}
\def\<#1>{\langle#1\rangle}
\def\cotg{\mathop{\rm cotg}\nolimits}
\def\Map{\mathop{\rm Map}\nolimits}
\def\wt{\widetilde}
\def\const{\hbox{const}}
\def\grad{\mathop{\rm grad}\nolimits}
\def\Div{\mathop{\rm div}\nolimits}
\def\braket#1#2{\langle#1|#2\rangle}
\def\Erf{\mathop{\rm Erf}\nolimits}
\def\matriz#1#2{\left( \begin{array}{#1} #2 \end{array}\right) }
\def\Eq#1{{\begin{equation} #1 \end{equation}}}
\def\deter#1#2{\left| \begin{array}{#1} #2 \end{array}\right| }
\def\pd#1#2{\frac{\partial#1}{\partial#2}}
\def\til{\tilde}

\def\la#1{\lambda_{#1}}
\def\teet#1#2{\theta [\eta _{#1}] (#2)}
\def\tede#1{\theta [\delta](#1)}
\def\N{{\frak N}}
\def\GR{{\cal G}}
\def\Wei{\wp}

\def\frac#1#2{{#1\over#2}} \def\pd#1#2{\frac{\partial#1}{\partial#2}}
                                                %  una derivada parcial
\def\matrdos#1#2#3#4{\left(\begin{matrix}#1 & #2 \cr          %para matrices 2X2
                                 #3 & #4 \cr\end{matrix}\right)}

\newtheorem{teor}{Teorema}[section]
\newtheorem{cor}{Corolario}[section]
\newtheorem{prop}{Proposici\'on}[section]
\newtheorem{note}[prop]{Note}
\newtheorem{definicion}{Definici\'on}[section]
\newtheorem{lema}{Lema}[section]
\theoremstyle{plain}
\newtheorem{theorem}{Theorem}
\newtheorem{corollary}{Corollary}
\newtheorem{proposition}{Proposition}
\newtheorem{definition}{Definition}
\newtheorem{lemma}{Lemma}

\def\Eq#1{{\begin{equation} #1 \end{equation}}}
\def\R{\Bbb R}
\def\C{\Bbb C}
\def\Z{\Bbb Z}
\def\mp#1{\marginpar{#1}}

\def\la#1{\lambda_{#1}}
\def\teet#1#2{\theta [\eta _{#1}] (#2)}
\def\tede#1{\theta [\delta](#1)}
\def\N{{\frak N}}
\def\Wei{\wp}
\def\Hil{{\cal H}}

\font\frak=eufm10 scaled\magstep1

\def\bra#1{\langle#1|}
\def\ket#1{|#1\rangle}
\def\goth #1{\hbox{{\frak #1}}}
\def\<#1>{\langle#1\rangle}
\def\cotg{\mathop{\rm cotg}\nolimits}
\def\cotanh{\mathop{\rm cotanh}\nolimits}
\def\arctanh{\mathop{\rm arctanh}\nolimits}
\def\wt{\widetilde}
\def\const{\hbox{const}}
\def\grad{\mathop{\rm grad}\nolimits}
\def\Div{\mathop{\rm div}\nolimits}
\def\braket#1#2{\langle#1|#2\rangle}
\def\Erf{\mathop{\rm Erf}\nolimits}

\centerline{\Large \bf Quantum Lie systems}
\vskip 0.75cm
\centerline{\Large \bf  and integrability conditions}
\vskip 0.75cm

\centerline{ Jos\'e F. Cari\~nena$^{\dagger}$ and Javier de Lucas$^{\dagger\ddagger}$}
\vskip 0.5cm

\centerline{$^{\dagger}$Departamento de F\'{\i}sica Te\'orica and IUMA, Universidad de Zaragoza,}
\medskip
\centerline{50009 Zaragoza, Spain.}
\medskip
\centerline{$^{\ddagger}$Institute of Mathematics of the Polish Academy of Science,}
\medskip
\centerline{P.O. Box 00-956, Warszawa, Poland.}
\medskip

\vskip 1cm

\begin{abstract}
 The theory of Lie systems has recently been applied to Quantum Mechanics and additionally
 some integrability conditions for Lie systems of differential
equations have also recently been analysed from a geometric perspective.
 In this paper we use both developments to obtain a geometric theory of integrability in Quantum Mechanics and we use it to provide a series of non-trivial
 integrable quantum mechanical models and to recover some known results from our
 unifying point of view.
\end{abstract}
\section{Introduction.}

\qquad Some recent papers have been devoted to apply the theory of Lie systems
\cite{LS,CGM07,PW,NI1} to Quantum Mechanics \cite{CLR07c,CarRam05b}. As a result it
has been proved that such a the theory can be used to treat some
types of Schr\"odinger equations, the so-called quantum Lie systems, to obtain
exact solutions, $t$-evolution operators, etc. One of the fundamental
properties found is that quantum Lie systems can be investigated by means of
equations in a Lie group. Through this equation we can analyse the properties
of the associated Schr\"odinger equation, i.e. the type of the associated Lie
group
allows us to know whether a Schr\"odinger equation can be integrated \cite{CLR07c}.

There are many papers devoted to study integrability of Lie systems,  for
instance the particular case of Riccati equations
\cite{CarRamGra,CRL07b}. It has been shown
in these papers that integrability conditions for Lie systems, in the
particular case of Riccati equations, appear as related to some transformation
properties of the associated equations in $SL(2,\mathbb{R})$. It was also shown in a recent work \cite{CRL07e} that the same procedure used to investigate Riccati equations can be applied to deal with any Lie system.

In the case of a quantum Lie system, the corresponding
Schr\"odinger equation can be associated with an equation in a Lie
group \cite{CLR07c}. The transformation properties of such an
equation  were
  investigated in the theory of integrability of Lie systems and they can be used in
  a quantum Lie system to study integrability conditions. All the
  results
 obtained in \cite{CRL07e} can be straightforwardly translated to the quantum
 framework and some non-trivial integrable models can be obtained.
The aim of this paper is to show how we can apply the theory of integrability
of Lie systems to quantum Lie systems and to give some applications.

The practical importance of this method is to be emphasised. It enables dealing with  non-trivially  integrable  $t$-dependent Schr\"odinger
equations. This fact allows us to investigate physical models by means of
non-trivial exact solutions. It also provides a procedure to avoid numerical
methods to study certain Schr\"odinger equations because when our methods can be applied, numerical methods are frequently not
necessary
and exact results can be used to test the efficiency and accuracy of
different approximation methods.

More specifically, in this paper we treat $t$-dependent spin Hamiltonians.
This kind of Hamiltonians appear broadly in Physics
\cite{AA87,Bl07,ANPT07,YLL97,ZY94} and, in particular, in the application of
the adiabatic approximation to the study of Berry phases
\cite{Sc37,PR08,MS04,XL06}. In this topic, some exact solutions of certain
non-trivial $t$-dependent spin Hamiltonians have been lately used to calculate
geometric phases exactly and through the adiabatic approximation, getting in
this way a method to check out the validity of such 
an approximation and solving some possible inconsistencies that have been lately pointed out. Furthermore, some works have recently been devoted to provide new integrable $t$-dependent spin Hamiltonians \cite{KN09PI,KN09PII,FR97}. In this paper, we explain why some of the $t$-dependent Hamiltonians appearing in all these works are integrable and we provide a method to get new integrable Hamiltonians in this and any other field where quantum Lie systems appear.

The organisation of the paper is as follows.  Section 2 is devoted to review
some properties of the Differential Geometry of infinite-dimensional manifolds
in order to analyse  in Section 3 the  theory of Lie systems and Quantum 
Mechanics. Section 4 describes the spin
Hamiltonian which appears broadly in many fields in Physics and  which gives rise
to the usually known as
Schr\"odinger--Pauli equation  \cite{CGM01} and
we show in this Section that the Schr\"odinger equation corresponding to this
Hamiltonian
 is a quantum Lie system. In Sections 5 and 6 the theory of integrability conditions
 developed in \cite{CRL07b, CRL07e} is applied to the Schr\"odinger equations studied in
 Section 4. In Section 7 we obtain some integrability conditions and in Section
 8  some applications to Physics are given.

\section{Differential geometry in Hilbert spaces.}\label{SLSQM}

\qquad In order to provide the basic knowledge to develop the main results of the paper, in this and next section we report some known results on the Differential Geometry in infinite-dimensional manifolds and quantum Lie systems. We also detail some results about quantum Lie systems which were not fully explained in previous works. For further details one can consult \cite{CLR07c, CarRam05b, KM97}.

As far as Quantum Mechanics is concerned, the separable complex
Hilbert space of states $\cal H$ can be seen as a
(infinite-dimensional) real manifold admitting a global chart
\cite{BCG}. Infinite-dimensional manifolds do not admit many of the classical results of the geometric
theory of finite-dimensional manifolds, e.g. in the most general
case and given an open $U\subset\mathcal{H}$, there is not a
one-to-one correspondence between derivations on
$C^{\infty}(U,\mathbb{R})$ and sections of the tangent bundle  $TU$. Therefore, some explanations must be done before dealing
with such manifolds.

On one hand, given a point $\phi\in\mathcal{H}$, a {\it kinematic tangent
  vector} with foot point $\phi$ is a pair $(\phi,\psi)$ with
$\psi\in\mathcal{H}$. We call $T_\phi\mathcal{H}$ the space of all kinematic
tangent vectors with foot point $\phi$. It consists of all derivatives $\dot c(0)$ at $\phi$ of smooth curves $c:\mathbb{R}\rightarrow\mathcal{H}$ with $c(0)=\phi$. This fact explains the name of kinematic.

From the concept of kinematic tangent vector we can provide the definition of
smooth kinematic vector fields as follows: A {\it smooth kinematic vector
  field} is an element $X\in \mathfrak{X}(\mathcal{H})\equiv \Gamma({\pi})$,
with $T\mathcal{H}$ the so-called {\it kinematic tangent bundle} and $\pi:{\rm
  T}\mathcal{H}\rightarrow\mathcal{H}$ the projection of this bundle. In
another way, we define a {\it kinematic vector field} $X$ as a map
$X:\mathcal{H}\rightarrow {\rm T}\mathcal{H}$ such that $\pi\circ X={\rm Id}_\mathcal{H}$. Given a $\psi\in\mathcal{H}$ we will denote from now on $X(\psi)=(\psi,X_{\psi})$ with $X_\psi$ the value of $X(\psi)$ in $T_\psi\mathcal{H}$.

In similarity with the Differential Geometry in finite-dimensional manifolds,
we say that a kinematic vector field $X$ on $\mathcal{H}$ has a local flow on
an open subset  $U\subset\mathcal{H}$ if there exists a map $Fl^X:\mathbb{R}\times U\rightarrow\mathcal{H}$ such that $Fl(0,\psi)=\psi$ for all $\psi\in U$ and
$$
X_{\psi}=\left.\frac{d}{dt}\right|_{t=0}Fl^X(t,\psi)=\left.\frac{d}{dt}\right|_{t=0}Fl^X_t(\psi),
$$
with $Fl^X_t(\psi)=Fl^X(t,x)$.

The definition of the Lie bracket at a point in a infinite
dimensional manifold for two kinematic vector fields $X,Y$
admitting local flows $Fl^X_t$ and $Fl^ Y_t$ coincides with the
known formula used in finite-dimensional Differential Geometry and
it reads
\begin{equation}\label{LieBrack}
[[X,Y]]_\psi=\left.\frac 12 \frac{d^2}{dt^2}\right|_{t=0}(Fl^Y_{-t}\circ Fl^X_{-t}\circ Fl^Y_{t}\circ F^X_{t}(\psi)).
\end{equation}

Let us use all these mathematical concepts to study Quantum Mechanics as a
geometric theory. Note that the Abelian translation group on $\mathcal{H}$ provides us with an
identification of  the tangent space $T_\phi\cal H$ at any point $\phi\in
\cal H$  with  $\cal H$ itself. Furthermore, through such an identification of $\cal H$ with $T_\phi\cal H$ at any $\phi\in \mathcal{H}$ a continuous
kinematic vector field is just a continuous
map $X\colon \cal H\to \cal H$, %�psi\mapsto X_\psi$.

Starting with a bounded $\mathbb{C}$-linear operator $A$ on $\mathcal{H}$, we
can define
the kinematic vector field $X^A$  by
$$
X^A_\psi=A\psi\in\mathcal{H}\simeq T_\psi\mathcal{H}.$$ 

Usually, operators
in Quantum Mechanics are neither continuous nor defined on the whole space
$\cal H$. The most relevant case is when $A$ is a skew-self-adjoint operator of the form $A=-i\, H$. The reason is that $\cal H$ can be endowed with
a natural (strongly) symplectic structure, and then such skew-self-adjoint
operators are singled out as the linear vector fields that are
Hamiltonian. The integral curves of such a Hamiltonian vector field
$X^A$ are the solutions of the corresponding Schr\"odinger equation
\cite{CLR07c,BCG}. Even when $A$ is not bounded, if $A$ is skew-self-adjoint it must be
densely defined and, by Stone's Theorem, its integral curves are strongly continuous and defined in all $\mathcal{H}$.

Additionally, these kinematic vector fields related to skew-self-adjoint operators admit local flows, i.e. any skew-self-adjoint operator $A$ has a local flow
\begin{equation}\label{Flow}
Fl^A_t(\psi)={\rm exp}(tA)(\psi)\quad {\rm as} \quad \frac{d}{dt}Fl^ A_t(\psi)=A{\rm exp}(tA)(\psi)=A(Fl^A_t(\psi)).
\end{equation}

We remark that given two constants $\lambda, \mu\in\mathbb{R}$ and skew-self-adjoint operators $A$ and $B$ we get that
$X^{\lambda A+\mu B}=\lambda X^A+\mu X^B$. Moreover,  skew-self-adjoint
operators considered as vector fields are
fundamental vector fields relative to the usual action of the unitary group
$U(\mathcal{H})$ on the Hilbert space $\mathcal{H}$.

Let us now investigate the Lie bracket of two kinematic vector fields $X^A$ and $X^ B$ associated with skew-self-adjoint operators $A$ and $B$. Indeed, taking into account (\ref{LieBrack}) and (\ref{Flow}), we get
\begin{equation*}\begin{aligned}
\left[\left[X^A,X^B\right]\right]_\psi&=\frac 12\frac{d^2}{dt^2}\bigg|_{t=0}\exp\left(-tB\right)\exp\left(-tA\right)\exp\left(tB\right)\exp\left(tA\right)(\psi)\\
&=\frac 12\frac{d^2}{dt^2}\bigg|_{t=0}\left(\sum_{n_1=0}^{\infty}\frac{(-tB)^{n_1}}{n_1!}\right)
\left(\sum_{n_2=0}^{\infty}\frac{(-tA)^{n_2}}{n_2!}\right)\\&
\left(\sum_{n_3=0}^{\infty}\frac{(tB)^{n_3}}{n_3!}\right)
\left(\sum_{n_4=0}^{\infty}\frac{(tA)^{n_4}}{n_4!}\right)(\psi)\\
&=\frac 12\frac{d^2}{dt^2}\bigg|_{t=0}\left(-t^2AB+t^2BA\right)(\psi)\\
&=\frac 12\frac{d^2}{dt^2}\bigg|_{t=0}\left(t^2[B,A]\right)(\psi)=[B,A](\psi),
\end{aligned}
\end{equation*}
from where we obtain that
\begin{equation}\label{FR}
[[X^A,X^B]]=-X^{[A,B]}.
\end{equation}

\section{Quantum Lie systems.}
\indent

Now the theory of Lie systems applies to the case in which a $t$-dependent
Hamiltonian can be written as a linear combination with $t$-dependent
real coefficients of some self-adjoint operators,
 \begin{equation}\label{LieHamiltonian}
 H(t)=\sum_{k=1}^rb_k(t)H_k\,,
\end{equation}
where the operators
$iH_k$ close on
a real finite-dimensional Lie algebra $V$ under the commutator of operators, i.e.
\begin{equation}
  [iH_j,iH_k]=\sum_{l=1}^rc_{jkl}\ iH_l,\qquad
  c_{jkl}\in\mathbb{R},\qquad j,k=1,\ldots,r.
\label{algebH}
\end{equation}
We call $V$ a {\it quantum Vessiot-Guldberg Lie algebra} and we say that $H(t)$ is a {\it quantum Lie system}.

A quantum Lie system $H(t)$ determines a Schr\"{o}dinger equation of the form
\begin{equation}\label{SchLie}
 \frac{d\psi}{dt}=-iH(t)\psi=-\sum_{k=1}^rb_k(t)iH_k\psi,
\end{equation}
describing the integral curves for the $t$-dependent kinematic vector field on $\mathcal{H}$ given by
$$X(t)=\sum_{k=1}^r b_k(t)X_k,$$
with $X_k$ the vector fields associated with the skew-self-adjoint operators ${-iH_k}$. In view of the relation (\ref{FR}) and the commutation relations (\ref{algebH}), we obtain
\begin{equation}
  [[X_j,X_k]]=\sum_{l=1}^rc_{jkl}X_l,\qquad j,k=1,\ldots,r,
\end{equation}
and we can get the constants of structure of the vector fields
$X_k$ by means of the commutators of the operators $iH_k$. The
linear combinations of the vector fields $X_k$, with
$k=1,\ldots,r$, span a real finite dimensional Lie algebra
isomorphic to $V$. On one hand, whether $\mathcal{H}$ is a
finite-dimensional manifold the Schr\"odinger equation
(\ref{SchLie}) is a Lie system. On the other hand, if
$\mathcal{H}$ were an infinite-dimensional manifold, such a
Schr\"odinger equation is the infinite-dimensional analogue to a
Lie system.

Now we can choose a basis $\{{\rm a}_k\,|\,k=1,\ldots,r\}$ of an abstract Lie
algebra $\goth{g}$ isomorphic to $V$ such that the Lie
brackets of the elements ${\rm a}_k$ of this Lie algebra, denoted by
$[\cdot,\cdot]$, satisfy
\begin{equation}
[{\rm a}_j,{\rm a}_k]=\sum_{l=1}^rc_{jkl}  {\rm a}_l\,,\qquad
  c_{jkl}\in\mathbb{R}.\label{Liealgdef}
\end{equation}
In this case one can show that there exists an unitary action
$\Phi:G\times\mathcal{H}\rightarrow\mathcal{H}$ of a Lie group
$G\subset U(\mathcal{H})$ with algebra $\LG$ on the Hilbert space
$\mathcal{H}$ such that the elements of the basis
 $\{{\rm a}_k\mid k=1,\ldots,r\}$ satisfying (\ref{Liealgdef}) hold
$$
\frac{d}{dt}\bigg|_{t=0}\Phi(\exp(-t{\rm a}_k),\psi)=(X_k)_\psi,
$$
i.e. the fundamental vector field associated with ${\rm a}_k$ is
the kinematic vector field $X_k$ related to the skew-self-adjoint
operator $-iH_k$.

 Then, solving the Schr\"odinger equation for the quantum Lie system $H(t)$ reduces to solving an equation in $G$ given by
$$
R_{g^{-1}*}\dot g=-\sum_{k=1}^rb_k(t){\rm a}_k\equiv {\rm a}(t),\qquad g(0)=e.
$$
Once that this equation has
been  solved the solution of the Schr\"odinger equation with initial condition $\psi(0)$ is $\psi(t)=\Phi(g(t),\psi(0))$, see \cite{CLR07c}.

\section{Spin Hamiltonians.}
\indent 

In this section we investigate a particular quantum mechanical system whose
dynamics is given by the Schr\"odinger--Pauli equation  \cite{CGM01}. We first prove that
this Hamiltonian is a quantum Lie system and  in a later Section we
 apply the theory of the integrability of Lie systems to such a system.

The system which we study is described by the $t$-dependent Hamiltonian
$$
H(t)=B_x(t)S_x+B_y(t)S_y+B_z(t)S_z,
$$
with $S_x, S_y$ and $S_z$ being the spin operators. Let us denote $S_1=S_x$, $S_2=S_y$ and $S_3=S_z$, then the  $t$-dependent Hamiltonian $H(t)$ is a quantum
Lie system,
because the spin operators are such that
\begin{equation}\label{ConmV}
[iS_j,iS_k]=-\sum_{l=1}^3\, \epsilon_{jkl}\,iS_l,\qquad j,k=1,2,3,
\end{equation}
with $\epsilon_{jkl}$ being the components of the fully skew-symmetric
 Levi-Civita tensor. The Schr\"odinger equation corresponding to this $t$-dependent Hamiltonian is
\begin{equation}\label{SE}
\frac{d\psi}{dt}=-iB_x(t)S_x(\psi)-iB_y(t)S_y(\psi)-iB_z(t)S_z(\psi),
\end{equation}
that can be seen as the differential equation for the determination of the integral curves for the $t$-dependent vector field in the (maybe-infinite dimensional) Hilbert space $\mathcal{H}$ given by
$$
X(t)=B_x(t)X_1+B_y(t)X_2+B_z(t)X_3,
$$
with
$$
({X_1})_\psi=-iS_x(\psi),\quad ({X_2})_\psi=-iS_y(\psi),\quad ({X_3})_\psi=-iS_z(\psi).
$$
The $t$-dependent vector field $X$ can be written as a linear combination
$X(t)={\displaystyle\sum_{k=1}^3} b_k(t)X_k$ of the vector fields $X_k$
with $b_1(t)=B_x(t)$, $b_2(t)=B_y(t)$ and $b_3(t)=B_z(t)$
and therefore  our Schr\"odinger equation is a Lie system related to a quantum Vessiot-Guldberg Lie algebra isomorphic to $\goth{su}(2)$.

A basis of $\goth{su}(2)$ such that their fundamental vector fields are the $X_k$ is given by the following
skew-selfadjoint $2\times 2$ matrices
\begin{equation}
{\rm a}_1\equiv\frac 12
\left(\begin{array}{cc}
0 & i\\
i & 0\\
\end{array}\right),\qquad
{\rm a}_2\equiv\frac 12
\left(\begin{array}{cc}
0 & 1\\
-1 & 0\\
\end{array}\right),\qquad
{\rm a}_3\equiv\frac 12
\left(\begin{array}{cc}
i & 0\\
0 & -i\\
\end{array}\right).
\end{equation}
These matrices satisfy the commutation relations
$$
[{\rm a}_j,{\rm a}_k]=-\sum_{l=1}^3\epsilon_{jkl}{\rm a}_l,\qquad j,k=1,2,3,
$$
which are similar to (\ref{ConmV}). Hence, we can define an action $\Phi:SU(2)\times\mathcal{H}\rightarrow\mathcal{H}$ such that
$$
\Phi(\exp(c_k{\rm a}_k),\psi)=\exp(c_k iH_k)(\psi),\qquad k=1,2,3,
$$
for any real constants $c_1, c_2$ and $c_3$. Moreover,
$$
\frac{d}{dt}\bigg|_{t=0}\Phi(\exp(-it{\rm a}_k),\psi)=\frac{d}{dt}\bigg|_{t=0}\exp(-itH_k)(\psi)=-iH_k(\psi)=({X_k})_\psi,
$$
getting that $X_k$ is the fundamental vector field associated with
${\rm a}_k$. Thus, the equation in $SU(2)$ related by means of
$\Phi$ to the Schr\"odinger equation (\ref{SE}) is
\begin{equation}\label{EQG}
R_{g^{-1}*g}\frac{dg}{dt}=-\sum_{k=1}^3b_k(t){\rm a}_k\equiv {\rm a}(t)\in\goth{su}(2),\qquad g(0)=e.
\end{equation}
It was shown in \cite{CLR07c} that the group $\mathcal{G}$ of curves in the group of a Lie system, in this case
$\mathcal{G}={\rm Map}(\R,SU(2))$, acts on the set of Lie systems associated with an
 equation in the Lie group $G$ in such a way that, in a similar way to what happened in \cite{CarRamGra}, a curve $\bar g\in\mathcal{G}$ transforms the initial equation (\ref{EQG}) into the new one characterised by the curve
\begin{equation}\label{trans}
{\rm a}'(t)\equiv -\Ad(\bar g)\left( \sum_{k=1}^3b_k(t){\rm a}_k\right)
+R_{\bar g^{-1}*\bar g}\frac{d\bar g}{dt}=-\sum_{k=1}^3b'_k(t){\rm a}_k.
\end{equation}
Once again this new equation is related to a new Schr\"odinger equation in $\mathcal{H}$ determined by a new Hamiltonian
$$
H'(t)=\sum_{k=1}^3b'_k(t)S_k\,.
$$
Additionally, it can be seen \cite{CLR07c} that the curve $\bar g(t)$ in $SU(2)$ induces a $t$-dependent unitary transformation $\bar U(t)$ on $\mathcal{H}$ transforming the initial $t$-dependent Hamiltonian $H(t)$ into $H'(t)$.

To sum up, the theory of Lie systems reduces the problem of
solving Schr\"odinger equations related to spin Hamiltonians
$H(t)$ to solve certain equations in the Lie group $SU(2)$. Then,
the transformation properties of the equations in $SU(2)$
describe the transformation properties of $H(t)$ by means of
certain $t$-dependent unitary transformations described by curves
in $SU(2)$.

Note also that the theory here developed for spin Hamiltonians can be straightforwardly applied to any other quantum Lie system. The only difference is that other quantum Lie system $H(t)$ may be related to other Lie group $G$. Anyway, the procedure described before can be applied again changing $SU(2)$ by the new Lie group $G$.

\section{Lie structure of an equation of transformation of Lie systems.}
\indent 

Our aim in this section is to prove that  the curves in $SU(2)$
relating the equations defined by two curves ${\rm a}(t)$ and ${\rm a}'(t)$ in
$T_ISU(2)\simeq\goth{su}(2)$, respectively, can be found as solutions of
 a Lie system of differential equations.

Recall that   the matrices of $SU(2)$ are of the form
\begin{equation}\label{parametrizacion}
\bar g=\left(
\begin{array}{cc}
a & b\\
-b^* & a^*
\end{array}
\right),\qquad a,b\in \mathbb{C},
\end{equation}
with $|a|^2+|b|^2=1$ and the elements of $\goth{su}(2)$ are traceless
skew-Hermitian matrices and therefore
 real linear combinations of the basis $\{{\rm a}_i\mid i=1,2,3\}$. Then the transformation rule (\ref{trans}), which can be considered as an equation in $SU(2)$ for the curve $\bar g$ relating to equations in $SU(2)$ determined
 by the curves ${\rm a}(t)$ and ${\rm a}'(t)$ in $\mathfrak{su}(2)$, becomes a matrix equation that can be written
\begin{equation}\label{trasRule}
\frac{d\bar g}{dt}\bar g^{-1}=-\sum_{k=1}^3b'_k(t){\rm a}_k+\sum_{k=1}^3b_k(t)\bar g{\rm a}_k\bar g^{-1}.
\end{equation}
Multiplying on the right by $\bar g$ both sides of this equation  we get
\begin{equation}\label{eqtran}
\frac{d\bar g}{dt}=-\sum_{k=1}^3b'_k(t){\rm a}_k\bar g+\sum_{k=1}^3b_k(t)\bar g{\rm a}_k\,.
\end{equation}
Considering a  reparametrisation of the $t$-dependent coefficients of $\bar g$
\begin{equation*}
\begin{aligned}
a(t)&=x_1(t)+i\,y_1(t),\\
b(t)&=x_2(t)+i\,y_2(t),
\end{aligned}
\end{equation*}
for real functions $x_j$ and $y_j$, with $j=1,2$, it is a straightforward computation to obtain that (\ref{eqtran}) is a linear
system of differential equations in the new variables $x_1,x_2,y_1$ and $y_2$
that can be written as follows:
\begin{equation}\label{FSys}
\matriz{c}{
\dot x_1\\
\dot x_2\\
\dot y_1\\
\dot y_2}
=\frac 12
\matriz{cccc}{
0&b'_2-b_2 &-b_3+b'_3&-b_1+b'_1\\
b_2-b'_2& 0 &-b_1-b'_1 &b_3+b'_3\\
b_3-b'_3& b'_1+b_1 & 0 & -b_2-b'_2\\
b_1-b'_1&-b_3-b'_3 &b_2+b'_2&0
}
\matriz{c}{
x_1\\
x_2\\
y_1\\
y_2}
\end{equation}

In order to describe the curves in $SU(2)$ relating two given
Schr\"odinger equations (\ref{SE}) with different $t$-dependent coefficients
as solutions of this system of  differential equations  we have to additionally
impose that $x_1^2+x_2^2+y_1^2+y_2^2=1$. Nevertheless, for the time being we can forget such a restriction because it is automatically implemented by means of a condition on the initial values of the variables $x_1,x_2,y_1$ and $y_2$ of the equation. Therefore we can deal with  the four variables
  in the preceding system of
 differential equations (\ref{FSys}) as being independent  ones. This linear
 system of  differential equations is a Lie system associated with a Lie algebra of vector fields $\mathfrak{gl}(4,\mathbb{R})$ but the solutions with initial condition related to a matrix in the subgroup $SU(2)$
   always remain in such a subgroup. In fact, consider the set of vector fields
\begin{eqnarray}\label{Vect}
N_1&=&\frac 12\left(-y_2\pd{}{x_1}-y_1\pd{}{x_2}+x_2\pd{}{y_1}+x_1\pd{}{y_2}\right),\cr
N_2&=&\frac 12\left(-x_2\pd{}{x_1}+x_1\pd{}{x_2}-y_2\pd{}{y_1}+y_1\pd{}{y_2}\right),\cr
N_3&=&\frac 12\left(-y_1\pd{}{x_1}+y_2\pd{}{x_2}+x_1\pd{}{y_1}-x_2\pd{}{y_2}\right),\cr
N'_1&=&\frac 12\left(y_2\pd{}{x_1}-y_1\pd{}{x_2}+x_2\pd{}{y_1}-x_1\pd{}{y_2}\right),\cr
N'_2&=&\frac 12\left(x_2\pd{}{x_1}-x_1\pd{}{x_2}-y_2\pd{}{y_1}+y_1\pd{}{y_2}\right),\cr
N'_3&=&\frac 12\left(y_1\pd{}{x_1}+y_2\pd{}{x_2}-x_1\pd{}{y_1}-x_2\pd{}{y_2}\right),\cr\nonumber
\end{eqnarray}
for which  the non-zero commutation relations are given by:
\begin{eqnarray}
\begin{aligned}
\left[N_1,N_2\right]=-N_3, \qquad &[N_2,N_3]=-N_1, \qquad &[N_3,N_1]=-N_2,\cr
[N'_1,N'_2]=-N'_3, \qquad &[N'_2, N'_3]=-N'_1,\qquad &[N'_3,  N'_1]=-N'_2\,.\nonumber
\end{aligned}
\end{eqnarray}

Note that $[N_j,N'_k]=0$, for $j,k=1,2,3$, and therefore
the system of linear differential equations (\ref{FSys}) is a Lie system in $\R^4$ associated with a Lie algebra of vector
fields isomorphic to $\LG\equiv\goth{su}(2)\oplus\goth{su}(2)$, i.e. the Lie
algebra decomposes into a direct sum of two Lie algebras isomorphic to
$\goth{su}(2)$, the first one is generated by $\{N_1,N_2,N_3\}$ and the second
one by $\{N'_1,N'_2,N'_3\}$.

If we  denote   $x\equiv\left(x_1,x_2,y_1,y_2\right)\in \R^4$,  the
system (\ref{FSys}) can be written as
 a system of differential equation in $\R^4$:
\begin{equation}\label{TRR4}
\frac{dx}{dt}=N(t,x),
\end{equation}
with $N(t)$ being the $t$-dependent vector field given by
\begin{equation*}
N(t)=\sum_{k=1}^3 \left(b_k(t)N_k+b'_k(t)N'_k\right).
\end{equation*}

The family of vector fields $\{N_1,N_2,N_3,N'_1,N'_2,N'_3\}$ generate an involutive distribution of rank
three in almost any point of $\R^4$ and consequently  there exists, at least locally, a first-integral for all
 the vector fields (\ref{Vect}). It can be verified that such a first-integral is globally defined and reads
 $I(x)=x_1^2+x_2^2+y_1^2+y_2^2$. Hence, given a solution $x(t)$ of the system (\ref{FS})
with $I(x(0))=x_1(0)^2+x_2(0)^2+y_1(0)^2+y_2(0)^2=1$,
then $ I(x(t))=1$  at any time $t$ and the solution $x(t)$ can be related to a curve in $SU(2)$. Therefore, we have found that
the curves in $SU(2)$
relating the  two different curves associated with two Schr\"odinger equations as in (\ref{SE}) can be
described by the solutions $x(t)$ of the system (\ref{TRR4}) with $I(x(0))=1$, and viceversa:

\begin{theorem}\label{ThMain} The curves in $SU(2)$  relating two equations in the group
  $SU(2)
$ characterised by the curves in $T_ISU(2)$
$$
{\rm a}'(t)=-\sum_{k=1}^3b'_k(t){\rm a}_k\,,
\qquad {\rm a}(t)=-\sum_{k=1}^3b_k(t){\rm a}_k$$
 are described by the solutions of the system
\begin{equation*}
\frac{dx}{dt}=N(t,x),
\end{equation*}
with
\begin{equation*}
N(t)=\sum_{k=1}^3\left(b_k(t)N_k+b'_k(t)N'_k\right),
\end{equation*}
and such that $I(x(0))=1$. This  is a Lie system related to a Lie algebra of
 vector fields isomorphic to
$\goth{su}(2)\oplus\goth{su}(2)$.
\end{theorem}
 As a consequence of the previous Theorem and the theorem of existence and uniqueness of solutions of differential equations, we get the next corollary.
\begin{corollary} \label{CorCur} Given two Schr\"odinger equations associated with
  curves ${\rm a}'(t)$ and ${\rm a}(t)$ in
$\goth{su}(2)$  there always exists  a curve in $SU(2)$ relating both systems.
\end{corollary}

Even if we know that given two equations in the Lie group $SU(2)$ there
exists always a transformation relating both which is a solution of the Lie
system (\ref{TRR4}),  in order to find such a curve one must explicitly solve
this system of  differential
equations.  This  is a Lie system related to a non-solvable Lie algebra, then it is
not  easy to find its solutions in general, e.g. it is not integrable by
quadratures.

From  this viewpoint, it is clear the existence of integrability conditions
that will be used in next Sections for obtaining exact solutions of some
physical problems in Quantum Mechanics.

\section{Integrability conditions for $SU(2)$ Schr\"odinger equations.}\label{GIC}
\indent 

We start  this section by reviewing some  results about integrability
conditions derived in \cite{CRL07e}
 from the perspective of the theory of Lie systems. The results given there are a little bit different from those
shown in this Section but the fundamental aspects remain the same. Indeed, the approach described along this Section can be applied to any Lie system in Classical and Quantum Mechanics. This fact clearly illustrates the importance of the method in order to obtain any kind of integrable Lie system.

As it was shown in the preceding Section, if the  curve $\bar g(t)\subset SU(2)$ transforms the equation in this Lie group defined by the  curve ${\rm a}(t)$ into
 another one characterised by ${\rm a}'(t)$ according to  the rule
 (\ref{trasRule}) and $g'(t)$
is a solution for the equation characterised by ${\rm a}'(t)$, then
$g(t)=\bar g^{-1}(t)g'(t)$ is a solution for the equation in $SU(2)$ characterised by ${\rm a}(t)$. Hence, the curve $\bar g(t)$ enables us to get the solution $g(t)$ of the equation in $SU(2)$ determined by ${\rm a}(t)$ from the solution $g'(t)$ for the equation determined by ${\rm a}'(t)$ and viceversa.

When  ${\rm a}'(t)$ lies in a solvable Lie subalgebra of $\goth{su}(2)$
we can obtain $g'(t)$ in many ways, e.g. by quadratures or by other  methods
as those used in  \cite{CarRamGra}. Then, once $g'(t)$ is obtained, the  knowledge  of the  curve $\bar g(t)$ transforming  the
 curve ${\rm a}(t)$ into ${\rm a}'(t)$ provides us the solution curve $g(t)$.

Therefore we begin with a curve
 ${\rm a}'(t)$ in a solvable Lie subalgebra of $\mathfrak{su}(2)$ and using
 (\ref{TRR4}), with curves in a restricted family of curves in $SU(2)$, we
relate it to other possible curves ${\rm a}(t)$, finding in this way
a family of equations in $SU(2)$, and thus spin Schr\"odinger equations in $\mathcal{H}$, that can be exactly solved.

Suppose we put some restrictions in the family of curves used in the system
differential equations (\ref{TRR4}), for instance we choose
$b=0$. As a consequence, there are instances for which
this
system of  differential equation does not admit any solution and it is not
possible to connect the  curves ${\rm a}(t)$ and ${\rm a}'(t)$
by a curve satisfying the assumed restrictions. Thus this gives rise to some
compatibility conditions for the existence of one of these special solutions, either
 algebraic and/or
 differential
ones, between the $t$-dependent  coefficients of ${\rm a}'(t)$ and ${\rm
  a}(t)$. It is shown later on that
such restrictions correspond to some integrability conditions that describes some exactly solvable models
proposed in the literature. So, our approach is useful to provide exactly integrable models found in the literature.

The two main ingredients to be taken into account in what follows are:

\begin{enumerate}
 \item {\it The equations which are characterised by a curve ${\rm a}'(t)$ for
which the solution can be obtained}.
We always suppose that ${\rm a}'(t)$ is related to a onedimensional Lie subalgebra of
$\goth{su}(2)$.

\item {\it The restriction on the set of curves considered as solutions of the
    equation (\ref{FS})}. In next sections we look for
solutions of (\ref{FS}) related to curves in a one-parameter
subset of $SU(2)$.
\end{enumerate}

Consider the next example of our theory: suppose we try to connect any ${\rm a}(t)$ with a final
family  ${\rm a}'(t)$ of curves of the form ${\rm a}'(t)=D(t)(c_1{\rm
  a}_1+c_2{\rm a}_2+c_3{\rm a}_3)$, with $c_i$ real numbers. In this way, the
system of differential equations (\ref{FS}) describing
 the curve $\bar g(t)\subset SU(2)$ which connects these curves reads:
\begin{equation}\label{Lie2}
\frac{dx}{dt}=\sum_{k=1}^3b_k(t)N_{k}(x)+D(t)
\sum_{k=1}^3c_k N'_k(x)=N(t,x).
\end{equation}
Now, as the vector field
\begin{equation*}
N'=\sum_{k=1}^3c_k N'_k,
\end{equation*}
is such that
\begin{equation*}
\left[N_k,N'\right]=0,\quad\quad  k=1,2,3,
\end{equation*}
the Lie system (\ref{Lie2}) is related to a Lie algebra of vector fields
isomorphic
to $\mathfrak{su}(2)\oplus \R$. As this Lie system has a non-solvable Vessiot-Guldberg Lie
algebra of vector fields, it is not integrable by quadratures and the solution
cannot be easily found in the general case. Nevertheless, it is worthy to
remark that (\ref{Lie2}) has  always a solution.

In this way we can consider some instances of (\ref{Lie2}) for which the
resulting system of differential equations can be integrated by quadratures. We can consider that $x$ is related to a one-parameter family of
elements of $SU(2)$. Such a restriction implies that (\ref{Lie2}) has not
always a solution because sometimes it is not possible   to connect ${\rm
  a}(t)$ and ${\rm a}'(t)$ by means of the chosen
one-parameter family. This fact  imposes differential and algebraic
restrictions to the initial $t$-dependent  functions $b_k$, with $k=1,2,3$. These restrictions
will describe previously known integrability conditions and other new ones. So,
we can develop the ideas of \cite{CRL07b, CRL07e} in the framework of Quantum Mechanics.
Moreover, from  this viewpoint we can find new integrability conditions that
 can be used to obtain exact solutions.

\section{Application of integrability conditions in a $SU(2)$ Schr\"odinger equation.}
\indent

In this Section we restrict ourselves to the case ${\rm a}'(t)=-D(t){\rm a}_3$,
i.e.
\begin{equation}
b'_1(t)=0,\qquad
b'_2(t)=0,\qquad
b'_3(t)=D(t).
\end{equation}
Hence, the system of differential equations (\ref{FSys}) describing the
 curves $\bar g$ relating a Schr\"odinger equation to
$H'(t)=D(t)S_z$
is
\begin{equation}\label{FS}
\matriz{c}{
\dot x_1\\
\dot x_2\\
\dot y_1\\
\dot y_2}
=\frac 12
\matriz{cccc}{
0&-b_2 &-b_3+D&-b_1\\
b_2& 0 &-b_1&b_3+D\\
b_3-D& b_1 & 0 & -b_2\\
b_1&-b_3-D &b_2&0
}
\matriz{c}{
x_1\\
x_2\\
y_1\\
y_2}.
\end{equation}
We see that according to the result of  Theorem \ref{ThMain} the $t$-dependent vector
field corresponding to such a  system of  differential equations can be written
 as a linear combination with $t$-dependent coefficients of the vector fields $N_1, N_2, N_3$ and $N_3'$:
$$
N(t)=\sum_{k=1}^3b_k(t)N_k+D(t)N'_3.
$$
Thus the system (\ref{FS}) is associated with a Lie algebra of vector fields
 isomorphic to $\goth{u}(1)\oplus
\goth{su}(2)$. This Lie algebra is simpler than the initial one  (\ref{FSys}),
but it is not solvable and the system  is as difficult to be solved as the
initial
 Schr\"odinger equation. Therefore in order to get exactly solvable cases we
 need to perform once again some kind of simplification, i.e. by means of some extra conditions on the variables. This means that the
obtained system of differential equation may  have no solution with the
 additional conditions imposed. The necessary and sufficient
conditions required to the initial and final $t$-dependent functions
in order to be able to  obtain a solution compatible with the assumed
restrictions
 give rise to integrability conditions for spin Hamiltonians.

Suppose for instance that we impose the solutions to be in the
one-parameter subset $A_\gamma\subset SU(2)$ given by
\begin{equation}\label{family}
A_\gamma=\left\{
\left(\begin{array}{cc}
         \cos\frac{\gamma}{2} & -e^{-bi}\sin\frac{\gamma}{2}\\
e^{bi}\sin\frac{\gamma}{2} &\cos\frac{\gamma}{2}
        \end{array}
 \right)\,\bigg |\,b\in [0,2\pi)
\right\},
\end{equation}
where $\gamma$ is a fixed real constant such that $\gamma\neq 2\pi n$, with $n\in\mathbb{Z}$, because in such a case $A_\gamma=\pm I$. In view of the definition of the sets $A_\gamma$ and in terms of the parametrisation (\ref{parametrizacion}) we have
\begin{equation*}
x_1=\cos\frac{\gamma}{2},\quad
y_1=0,\quad
x_2=-\sin\frac \gamma 2\cos b,\quad
y_2=\sin\frac \gamma 2\sin b.
\end{equation*}
The elements of $A_\gamma$ are matrices in $SU(2)$ and  the system of differential equations we obtain reads
\begin{equation}\label{sysInt}
\matriz{c}{
0\\
\dot x_2\\
0\\
\dot y_2}
=\frac 12
\matriz{cccc}{
0&-b_2 &-b_3+D&-b_1\\
b_2& 0 &-b_1&b_3+D\\
b_3-D& b_1 & 0 & -b_2\\
b_1&-b_3-D &b_2&0
}
\matriz{c}{
x_1\\
x_2\\
0\\
y_2}.
\end{equation}
and then we get as two integrability conditions
for the system (\ref{sysInt}):
\begin{equation}\label{Con}
\begin{aligned}
0&=-b_2x_2-b_1y_2,\\
0&=(b_3-D)x_1+b_1x_2-b_2y_2.
\end{aligned}
\end{equation}
We can write the components $(B_x(t),B_y(t),B_z(t))$ of the magnetic field in polar coordinates,
$$
\begin{aligned}
B_x(t)&=B(t)\sin\theta(t)\cos\phi(t),\\
B_y(t)&=B(t)\sin\theta(t)\sin\phi(t),\\
B_z(t)&=B(t)\cos\theta(t),
\end{aligned}
$$
with $\theta\in[0,\pi)$ and $\phi\in [0,2\pi)$.

The first algebraic integrability condition reads in polar coordinates as follows:
$$
B(t)\sin\theta(t)\sin\frac \gamma 2\left(\cos\phi(t)\sin b(t)-\sin\phi(t)\cos b(t)\right)=0
$$
and thus
$$
B(t)\sin\theta(t)\sin\frac\gamma 2\sin(b(t)-\phi(t))=0,
$$
from where we see  that $b(t)=\phi(t)$. In such a case the second algebraic integrability condition in (\ref{Con}) reduces to
\begin{equation*}
(B_z-D)\cos\frac \gamma 2-B\sin\frac \gamma 2\sin\theta=0
\end{equation*}
and then the $t$-dependent coefficient $D$ is
\begin{equation}\label{DFactor}
D=\frac{B}{\cos\frac \gamma 2}\cos\left(\frac \gamma 2+\theta\right).
\end{equation}
Finally we have to take into account the differential integrability condition
$$
\dot x_2=\frac 12\left(b_2\cos\frac \gamma 2+(b_3+D)\sin\frac \gamma 2\sin b\right),
$$
which after some algebraic manipulation leads to
$$
\dot \phi=\frac{B}2\left(\frac{\sin(\theta+\frac \gamma 2)}{\sin\frac \gamma 2}+\frac{\cos(\frac \gamma 2+\theta)}{\cos\frac \gamma 2}\right),
$$
and then
\begin{equation}\label{IntCond}
\dot \phi=B\,\frac{\sin(\theta+\gamma)}{\sin\gamma},
\end{equation}
which is a far larger set of integrable Hamiltonians than the one of the
exactly solvable Hamiltonians of this type
 found  in the literature. As a particular example, when $\theta$ and $B$ are
 constants
 we find
\begin{equation}\label{CI}
\dot \phi=B\,\frac{\sin(\theta+\gamma)}{\sin\gamma}\equiv \omega,
\end{equation}
and consequently,
$$
\phi=\omega t+\phi_0.
$$
In this way, we get that the $t$-dependent spin Hamiltonian $H(t)$ determined by the magnetic vector field
\[
{\bf B}(t)=B(\sin\theta\cos(\omega t),\sin\theta\sin(\omega t),\cos\theta)
\]
is integrable.

Another interesting integrable case is that given by
$\theta=\pi/2$, that is, the magnetic field moves in the $XZ$
plane, see \cite{KN09PII, Bl08}. In such a case, in view of the
integrability conditions (\ref{CI}), the angular frequency
$\dot\phi$ reads
$$
\dot\phi=-B\,{\rm cotan}\gamma=\omega.
$$

The last one of the most known integrable cases, that given
by a magnetic field in a fixed direction ${\bf
B}(t)=B(t)(\sin\theta\cos\phi,\sin\theta\sin\phi,\cos\theta)$,
satisfies the integrability condition (\ref{IntCond}) for
$\gamma=-\theta$.

Apart from the previous cases, the integrability condition (\ref{IntCond}) describes more, as far as we know new, integrable cases.
For instance, consider the case with $\theta$ fixed and $B$ non-constant. In this case, the corresponding $H(t)$ is integrable if
$$
\frac{\dot\phi}{B}=\frac{\sin(\theta+\gamma)}{\sin\gamma},
$$
that is, if we fix $\gamma=\pi/2$ we have that
$$
\omega=\dot \phi=-B(t)\cos\theta\Longrightarrow\phi(t)=-\cos\theta\int^tB(t')dt'.
$$

Furthermore, we can consider $\theta(t)=t$ and $B$ constant. In this case, if we fix $\gamma=-\pi/2$, we get that the $t$-dependent Hamiltonian $H(t)$ is integrable if the $\phi(t)$ for ${\bf B}(t)$ holds the condition
$$
\dot\phi=-B\cos\,t\Longrightarrow \phi(t)=-B\sin\,t.
$$

To sum up, we have shown that there exists a complete family of $t$-dependent
integrable spin Hamiltonians much broader than the until now known integrable
cases. It is 
also easy to verify if a $t$-dependent spin Hamiltonian holds the integrability condition (\ref{IntCond}) and then it can be integrated.
\section{Applications in Physics.}
\indent

In this section we use the results of the previous Section in order to solve a $t$-dependent spin Hamiltonian
\[
H(t)={\bf B}(t)\cdot{\bf S},
\]
which appears broadly Physics: that one characterised by the particular magnetic fields
\begin{equation}\label{MF}
{\bf B}(t)=B(\sin\theta\cos(\omega t),\sin\theta\sin(\omega t),\cos\theta),
\end{equation}
that is, magnetic fields with a constant modulus rotating along the $OZ$  axis
with
a  constant angular velocity $\omega$. Such Hamiltonians have been applied, for instance, to analyse the precession of a spin in a transverse  $t$-dependent magnetic field \cite{Sc37}, investigate the adiabatic approximation and the unitarity of the $t$-evolution operator through such an approximation \cite{PR08,MS04}, etc.

In the previous Section we showed that this $t$-dependent Hamiltonian is integrable. Indeed, the integrability condition (\ref{CI}) can be written as
\begin{equation}\label{equation}
\tan \gamma=\frac{\sin\theta}{\frac{\dot\phi}{B}-\cos\theta},
\end{equation}
where we recall that $\gamma$ must be a real constant. In the case of our
particular magnetic field (\ref{MF}) the angular frequency, $\omega=\dot\phi$,
the angle $\theta$ and  the modulus $B$ are constants. Therefore $\gamma$ is a properly defined constant, the integrability condition (\ref{CI}) holds and the value of $\gamma$ is given by equation (\ref{equation}) in terms of the parameters $B$, $\theta$ and $\omega$ describing the magnetic vector field.

We have already shown that whether $B(t)$ satisfies (\ref{CI}) then $H(t)$ is
integrable because it can be transformed by means of a $t$-dependent change of
variables in $\mathcal{H}$ induced by a curve $g(t)$ in the set $A_\gamma$ into
a straightforwardly integrable Schr\"odinger equation determined by a
$t$-dependent Hamiltonian $H'(t)=D(t)S_z$.  For the sake of simplicity let us
parametrise the elements of $A_\gamma$ in a new way. Consider that being
$\overrightarrow{\sigma}=(\sigma_1,\sigma_2,\sigma_3)$ and $\overrightarrow{n}\in\mathbb{R}$ with the matrices $\sigma_i$ the Pauli matrices $\sigma_x,\sigma_y$ and $\sigma_z$, we have
$$
e^{i\overrightarrow{\sigma}\cdot\overrightarrow{n}\phi}=I\cos\phi+i\overrightarrow{\sigma}\cdot\overrightarrow{n}\sin\phi.
$$
So, for $\overrightarrow{n}=(\alpha_1,\alpha_2,0)/\sqrt{\alpha_1^2+\sigma_2^2}$ with real constants $\alpha_1$, $\alpha_2$ and taking into account that ${\rm a}_1=i\sigma_x/2,{\rm a}_2=i\sigma_y/2$ and ${\rm a}_3=i\sigma_z/2$, we get that
\begin{equation}\label{para}
\exp(\alpha_1{\rm a}_1+\alpha_2{\rm a}_2)=\exp(i\frac{\delta}2\overrightarrow{\sigma}\overrightarrow{n})=
\left(\begin{array}{cc}
{\rm cos}\frac{\delta}{2}&-e^{-i\varphi}\sin\frac{\delta}{2}\\
e^{i\varphi}\sin\frac{\delta}{2}&\cos \frac{\delta}{2}
\end{array}\right)
\end{equation}
with $\delta=\sqrt{\alpha_1^2+\alpha_2^2}$ and $e^{-i\varphi}=(\alpha_1+i \alpha_2)/\sqrt{\alpha_1^2+\alpha_2^2}$. In terms of $\delta$ and $\varphi$ the variables $\alpha_1$ and $\alpha_2$ can be written $\alpha_1=\delta\sin\varphi$ and $\alpha_2=-\delta\cos\varphi$.
Hence, in view of (\ref{para}), we see that we can describe the elements of $A_\gamma$ as
\begin{equation}
\left(\begin{array}{cc}
         \cos\frac{\gamma}{2} & -e^{-bi}\sin\frac{\gamma}{2}\\
e^{bi}\sin\frac{\gamma}{2} &\cos\frac{\gamma}{2}
        \end{array}
 \right)=\exp(\gamma \sin b\,{\rm a}_1-\gamma \cos b\,{\rm a}_2),
\end{equation}
where $b$ and $\gamma$ are real constants. For magnetic vector fields (\ref{MF}), the $t$-dependent change of variables transforming the initial $H(t)$ into an integrable $H'(t)=D(t)S_z$ is determined by a curve in $A_\gamma$ with $\gamma$ determined by the equation (\ref{equation}) and $b(t)=\phi(t)$. Thus, such a curve in $A_\gamma$ has the form
\begin{equation}\label{tra}
t\mapsto \exp(\gamma \sin( \omega t)\,{\rm a}_1-\gamma \cos( \omega t)\,{\rm a}_2).
\end{equation}
We emphasise that the above $t$-dependent change of variables in $SU(2)$ transforms the equation in $SU(2)$ determined by the initial curve
$${\rm a}(t)=-B_x(t){\rm a}_1-B_y(t){\rm a}_2-B_z(t){\rm a}_3,
$$
into and a new equation in $SU(2)$ determined by a curve ${\rm
a}'(t)=-D(t){\rm a}_3$. Such a $t$-dependent transformation in
$SU(2)$ induces a $t$-dependent unitary change of variables in
$\mathcal{H}$ transforming the initial Schr\"odinger equation
determined by the $t$-dependent Hamiltonian $H(t)$, i.e.
$$
\frac{\partial\psi}{\partial t}=-iH(t)(\psi),
$$
into the new Schr\"odinger equation
\begin{equation}\label{finalSC}
\frac{\partial\psi'}{\partial t}=-iH'(t)(\psi')=-iD(t)S_z(\psi').
\end{equation}
being the relation between $\psi$ and $\psi'$ given by the corresponding $t$-dependent change of variables in $\mathcal{H}$ induced by the curve (\ref{tra}), i.e.
\begin{equation}\label{Change}
\psi'=\exp(\gamma \sin (\omega t)\,iS_x-\gamma \cos (\omega t)\,iS_y)\psi.
\end{equation}
In view of expression (\ref{DFactor}), we see that
$$D=B(\cos\theta-{\tan}\frac\gamma 2\sin\theta),$$
and from (\ref{equation}) ant taking into account that
$${\tan}\gamma=\frac{2{\tan}\frac \gamma 2}{1-{\tan}^2\frac\gamma 2}\Rightarrow {\tan}\frac\gamma 2=\frac {-1\pm\sqrt{1+{\tan}^2\gamma}}{{\tan}\gamma},
$$
we obtain
$$
{\tan}\frac\gamma
2=\frac{1}{\sin\theta}\left(-\frac{\omega}{B}+\cos\theta\pm\sqrt{\frac{\omega^2}{B^2}-2\frac\omega
B\cos\theta +1}\right).
$$
Using this result in the latter expression for $D$, such a
function becomes a constant which can be written in terms of the
variables $\theta$, and $B$ as
$$
D=\omega\pm\sqrt{\omega^2-2\omega B\cos\theta+B^2}.
$$
Thus the general solution $\psi'(t)$ for the Schr\"odinger
equation (\ref{finalSC}) with initial condition $\psi'(0)$ is
$$
\psi'(t)=\exp\left(-itDS_z \right)\psi'(0).
$$
And the solution for the initial Schr\"{o}dinger equation with initial
condition $\psi(0)$ can
 be obtained by inverting the $t$-dependent change of variables (\ref{Change}) to get
$$
\psi(t)=\exp\left(-i\gamma\sin \omega t \,S_x+i\gamma\cos \omega t\, S_y\right)\exp\left(-iD tS_z \right)\psi(0).
$$

\section{Conclusions and Outlook}

\indent
We have shown that some previous results of the theory of integrability conditions developed in \cite{CRL07e} are
straightforwardly applicable to a larger set than Riccati equations. Indeed, we
have
shown that these results can be applied to Schr\"odinger equations on infinite-dimensional manifolds in order to obtain
non-trivial
 exactly  solvable $t$-dependent quantum Hamiltonians that are used in many fields of Physics.

As a consequence of this  it is clear  that it is possible to generalise the
 procedure shown in \cite{CRL07e} to  a more general
 framework and this fact is worthy of  a deeper study.
\section*{Acknowledgements}

 Partial financial support by research projects MTM2006-10531 and E24/1 (DGA)
 are acknowledged. JdL also acknowledges
 a F.P.U. grant from  Ministerio de Educaci\'on y Ciencia.

\end{document}